\documentclass{article}
\usepackage{ijcai16}

\usepackage{times}
\usepackage{epsfig}
\usepackage{graphicx}
\usepackage{amsmath}
\usepackage{amssymb}

\usepackage{algorithm}
\usepackage{algorithmic}
\usepackage{multirow}
\usepackage{epsfig}
\usepackage{subfigure}
\usepackage{times}
\usepackage{graphicx}
\usepackage{amssymb}
\usepackage{float}

\begin{document}

\title{Supervised Matrix Factorization for Cross-Modality Hashing}
\author{Hong Liu$^{\dag\ddag}$, 
Rongrong Ji$^{\dag\ddag}$\thanks{Corresponding author.}, 
Yongjian Wu$^{\natural}$,
Gang Hua$^{\flat}$ \\
$^{\dag}$Fujian Key Laboratory of Sensing and Computing for Smart City, Xiamen University, 361005, China\\
$^{\ddag}$School of Information Science and Engineering, Xiamen University, 361005, China\\
$^{\natural}$BestImage, Tencent Technology (Shanghai) Co.,Ltd, China\\
$^{\flat}$Microsoft Research Asia, Beijing, China\\
{\texttt{lynnliuxmu@outlook.com, rrji@xmu.edu.cn, littlekenwu@tencent.com,}} \\
\texttt{ganghua@gmail.com} \\
}

\maketitle

\begin{abstract}
Matrix factorization has been recently utilized for the task of multi-modal hashing for cross-modality visual search, where basis functions are learned to map data from different modalities to the same Hamming embedding.  
%
In this paper, we propose a novel cross-modality hashing algorithm termed Supervised Matrix Factorization Hashing (SMFH)  which tackles the multi-modal hashing problem with a collective non-negative matrix factorization across the different modalities.
In particular, SMFH employs a well-designed binary code learning algorithm to preserve the similarities among multi-modal original features through a graph regularization. 
At the same time, semantic labels, when available, are incorporated into the learning procedure.
We conjecture that all these would facilitate to preserve the most relevant information during the binary quantization process, and hence improve the retrieval accuracy. 
We demonstrate the superior performance of SMFH on three cross-modality visual search benchmarks, \textit{i.e.}, the PASCAL-Sentence, Wiki, and NUS-WIDE, with quantitative comparison to various state-of-the-art methods  \cite{kumar2011learning,rastegari2013predictable,zhang2014large,ding2014collective}.
\end{abstract}

\section{Introduction}

Cross-modality retrieval has been a fundamental problem in several emerging applications including visual search, machine translation, and text mining \cite{bronstein2010data,masci2014multimodal,rasiwasia2010new,costa2014role}. In a typical scenario of cross-modality retrieval, a query comes from one modality, \textit{e.g.}, text, while the returned results come from another modality, \textit{e.g.}, image. To achieve this goal, a typical solution adopted  by most existing works is to embed data samples of different modalities into a common low-dimensional space. 
By doing so, both the query and returns can be well aligned to capture their cross-modality similarities for retrieval \cite{costa2014role,wang2014effective}.

Recently, both unsupervised and supervised hashing techniques have been investigated for cross-modality retrieval due to their prominent efficiency. For instance, Bronstein \emph{et al.} proposed a Cross-Modality Similarity Search Hashing (CMSSH) algorithm by using eigen-decomposition and boosting. Both Cross-View Hashing (CVH) \cite{kumar2011learning} and Inter-Media Hashing (IMH) \cite{song2013inter} extended  the classic Spectral Hashing approach \cite{weiss2009spectral} to the scenario of cross-modality retrieval. For another instance, Co-Regularized Hashing (CRH) \cite{zhen2012co} and Heterogeneous Translated Hashing (HTH) \cite{wei2014scalable} further deal with the cross-modality hashing under a co-regularized boosting framework. 

In \cite{rastegari2013predictable}, Predictable Dual-View Hashing (PDH) was proposed to learn the discriminative hash functions via a max-margin formulation with an iterative optimization algorithm. In \cite{ding2014collective}, Collective Matrix Factorization Hashing (CMFH) was proposed to formulate the joint learning of cross-modality binary codes as a collective matrix factorization problem. In \cite{zhang2014large}, Supervised Multi-modal Hashing (SMH) was proposed to integrate semantic labels to improve the performance of hash function learning in the respective modalities. 

While promising progress has been made, it remains as an open problem to capture the multi-modal similarities among data samples, as well as to preserve such similarities in a produced binary code (Hamming) space.
As mentioned before, collective factorization \cite{ding2014collective} and supervised hashing \cite{zhang2014large} have demonstrated outperformance on respective problems, \emph{a.k.a.}, for cross-modality retrieval and for single-modality binary code learning \cite{liu2015multiview,mukherjee2015nmf,lin2015semantics}.
The intuition of our work is to combine the merits of supervised hashing \cite{liu2012supervised,zhang2014large} over the state-of-the-art cross-modality retrieval schemes \cite{ding2014collective}.

However, the integration of both approaches towards supervised cross-modality hashing is not an easy task.
In one aspect, it is hard to optimize the discrete Hamming distances.
In the other aspect, the complexity of the existing matrix factorization hashing and supervised cross-modality hashing are very high, which are the square of the training set size \cite{zhang2014large,liu2015multiview,mukherjee2015nmf} and cannot be easily scaled up to massive training data.

\begin{figure*}[!t]\label{fig1}
\begin{center}
\includegraphics[height=0.35\linewidth, width=0.7\linewidth]{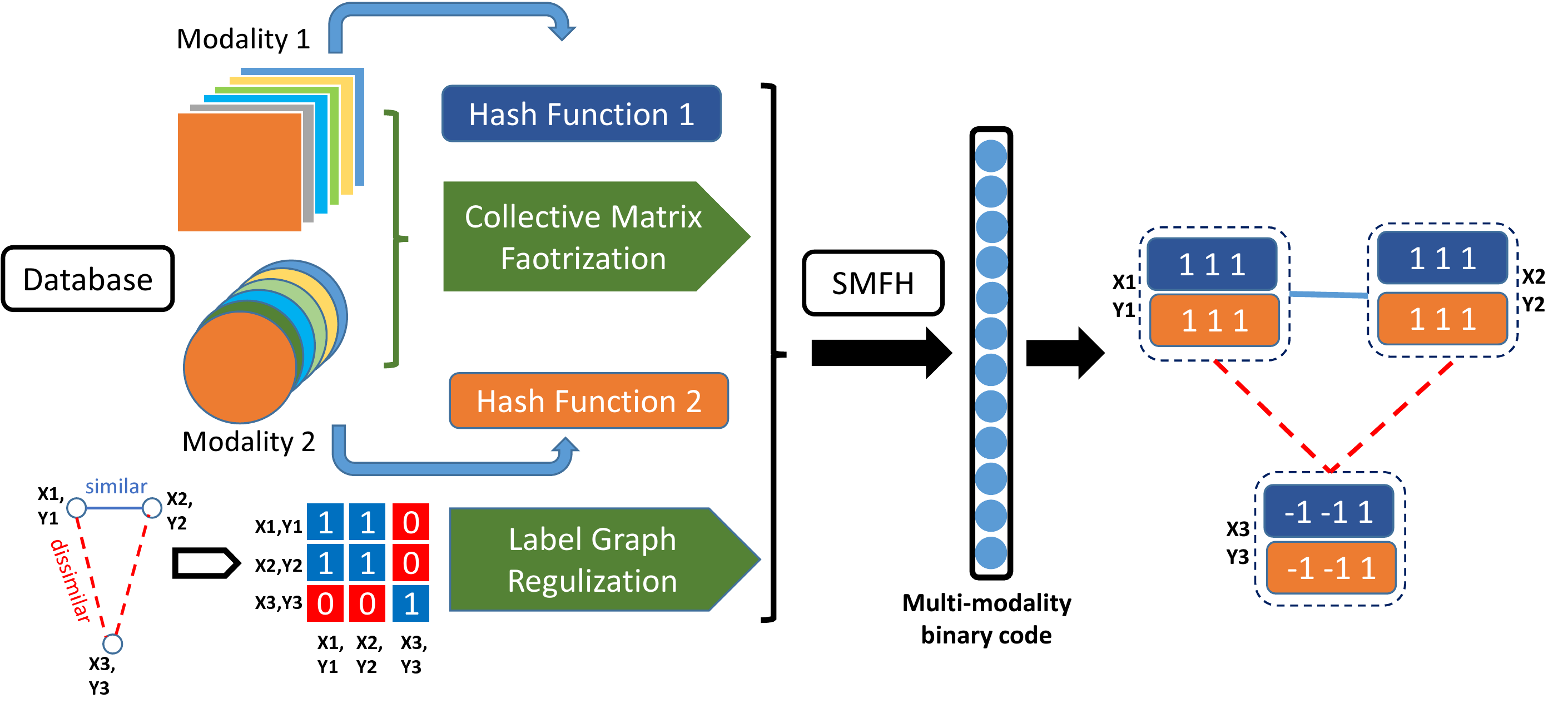}
\caption{ The Framework of our proposed Supervised Matrix Factorization Hashing (SMFH).}
\end{center}
\end{figure*}

In this paper, we propose a novel cross-modality hashing method, dubbed the name Supervised Matrix Factorization Hashing (SMFH), which addresses the above challenges under a graph-regularized, collective non-negative matrix factorization framework. 
The first contribution we make is a hybrid regularization method to model the hash function learning, which integrates graph regularization into the collective non-negative matrix factorization. 
The second contribution we make is a supervised collective non-negative matrix factorization scheme, which leverages semantic labels to refine the graph regularizer during the step of graph construction. 
It ensures the learned binary codes to preserve the semantic similarities among data within multiple modalities. 

Besides the two contributions mentioned above, we also propose an optimization algorithm to solve the objective function designed for SMFH, which works under an iterative updating procedure with stochastic sampling.
This strategy can reach a training time reduction of supervised cross-modality hashing. 
%
Fig. 1 shows the overall framework of the proposed SMFH scheme.
We conduct extensive experiments in cross-modality visual search, \textit{i.e.}, using text queries to retrieve relevant images and vice versa, on three widely used benchmarks including, PASCAL-Sentence, Wiki and NUS-WIDE. We demonstrate the superior performance of SMFH over a group of state-of-the-art cross-modality hashing methods including CVH \cite{kumar2011learning}, PDH \cite{rastegari2013predictable}, SMH \cite{zhang2014large}, and CMFH \cite{ding2014collective}.

The rest of this paper is organized as follows: in Section \ref{Sec3}, we present our SMFH approach in depth. Section \ref{Sec4} shows extensive experiments conducted on three benchmark datasets. Finally, we draw our conclusions in Section \ref{Sec5} and discuss the future work.

\section{Supervised Matrix Factorization Hashing}\label{Sec3}

In this section, we describe the proposed supervised cross-modality hashing algorithm.
Without loss of generality, we take bi-modal hashing for instance, which can be easily extended to the scenario of multi-modality hashing.

\subsection{Collective Factorization for Cross-Modality Hashing}

Non-negative Matrix Factorization (NMF) is a matrix decomposition algorithm that focuses on learning low-rank representation.
We define a non-negative data matrix $\mathbf{X}=[x_1,...,x_N]\in{\mathbb{R}^{d\times{N}}}$, where $N$ is the number of samples, $d$ is the feature dimension,  and $x_i$ is the $i$-th sample.
Non-negative matrix factorization aims at finding two non-negative factors 
$\mathbf{U} = [u_{1},...,u_{r}]\in{\mathbb{R}^{d\times{r}}}$
and
$\mathbf{Y} = [v_{1},...,v_{N}]\in{\mathbb{R}^{r\times{N}}}$, $r<< d$, 
whose product can approximate $\mathbf{X}$, \textit{i.e.},
\small{
\begin{equation}\label{Eq1}
\mathbf{X} \approx{\mathbf{U}\mathbf{Y}}.
\end{equation}}
The squared Frobenius norm of the difference between two matrices is commonly used as the cost function to measure the approximation quality \cite{cai2011graph,liu2013multi}, which is defined as: 
\small{
\begin{equation}\label{Eq2}
\min _{\mathbf{U},\mathbf{Y}}{\parallel \mathbf{X} - \mathbf{U}\mathbf{Y} \parallel_{F}^{2} },   \; s.t. \ \mathbf{U}\geqslant 0, \mathbf{Y}\geqslant 0 , 
\end{equation}}
where $\parallel \cdot  \parallel_{F}$ is the Frobenius norm of the matrix.

However,  when facing cross-modality data, it is expected that the matrix $\mathbf{Y}$ is a new multi-modality representation of  the data matrices $\mathbf{X}^1$ and $\mathbf{X}^2$ in a common low-dimensional space.
%
%
It aims to learn two base $\mathbf{U}^1\in{\mathbb{R}^{d_1\times{r}}}$ and $\mathbf{U}^2\in{\mathbb{R}^{d_2\times{r}}}$ to produce hash codes, which map bi-modal feature matrices into a $r$-dimensional binary code matrix $\mathbf{Y}\in{\{0,1\}^{r\times{N}}}$, where $r$ is the number of hash bits. Ideally, $\mathbf{Y}$ should reveal the hidden semantics shared by different modalities.

Correspondingly, the learning of the hash code is done via:
\small{
\begin{equation}\label{Eq3}
\min_{\mathbf{Y},\mathbf{U}^1,\mathbf{U}^2}\sum_{t=1}^{2} \lambda_t{\left \| \mathbf{X}^t-\mathbf{U}^t\mathbf{Y} \right \|_{F}^{2}},
\end{equation}}
where $\lambda_t$ is the weight coefficient of this modality satisfying $\sum\nolimits_{t=1}^{2}\lambda_t=1$.
Intuitively, for the $1$-st modal data $\mathbf{X}^1$, the matrix $\mathbf{U}^1$ and the matrix $\mathbf{Y}$ are learned by Eq.  (\ref{Eq3}), which is the same to $\mathbf{X}^2$.
$y_{ji} = 1$ means that the corresponding semantic representation $u^t_{\cdot{j}}$ is related to data $\mathbf{X}^t_{\cdot{i}}$ and vice versa.

\subsection{Learning Supervised Hash Function}

Towards learning supervised hash function, our goal is to preserve the semantic similarity among data points in the Hamming space. 
That says, the hash functions should enforce labeled data pair $l(e_i)=l(e_j)$ to have identical or similar binary codes, 
where $e_i=(x_i^1,x_i^2)$ is the $i$-th data sample on both modalities from the training set, and function $l(\cdot)$ returns the supervised labels of the data across different modalities.

Our task is to find a graph matrix that preserves the intrinsic geometric structure of the  similarity from bi-modal data.
To this end, we first construct an affinity graph to model the aforementioned semantic information.
This is done by calculating the pairwise similarity via the inner product among the semantic labels.

More specifically, the similarity between the $i$-th data sample and the $j$-th data sample is  defined as follow:
\small{
\begin{equation}\label{Eq4}
\mathbf{A}_{ij} = l(e_i)^T \cdot l(e_j).
\end{equation}}
Without loss of generality, we define a matrix $\mathbf{L_a}\in{\{0,1\}^{dl\times{N}}}$, where each column of $\mathbf{L_a}$ is the label representation vector of each data sample, and each row represents each sample's category\footnote{In this paper, we assume that $L_a$ is fully observed without missing labels. We can get the labels among points in many cases for missing labeled datas, \textit{i.e.} classification technology, which makes our assumption reasonable.}.  
Subsequently, the similarity matrix can be presented as $\mathbf{A}=\mathbf{L_a}^T{\cdot}\mathbf{L_a}$.

Then, our goal can be formulated via a spectral graph learning problem from the label similarity matrix  $\mathbf{A}$ as:
\small{
\begin{equation}\label{Eq5}
\min_{\mathbf{Y}}\frac{1}{2}\sum_{i,j=1}^{n}{\left \| \mathbf{Y}_{i} - \mathbf{Y}_{j} \right \|^{2}\mathbf{A}_{i,j}}= {Tr(\mathbf{Y}^{T}\mathbf{L}\mathbf{Y})},
\end{equation}}
where $\mathbf{L}$ is the Laplacian matrix for $\mathbf{A}$.

Given data in two modalities, \textit{a.k.a}, $\mathbf{X}^1 = \{x^1_{i}\in{\mathbb{R}^{d_1}}|i=1,...,N\}$, $\mathbf{X}^2 = \{x^2_{i}\in{\mathbb{R}^{d_2}}|i=1,..., N\}$, and their pairwise semantic similarity $\mathbf{A}$,
the goal of supervised hash function learning is to learn two basis matrices $\mathbf{U}^1$ and $\mathbf{U}^2$, together with the hash codes $\mathbf{Y}$ by the following objective function: 
\small{
\begin{equation}\label{Eq6}
\min_{\mathbf{Y},\mathbf{U}^1,\mathbf{U}^2} \sum_{t=1}^{2} \lambda_t{\left \| \mathbf{X}^t-\mathbf{U}^t\mathbf{Y} \right \|_{F}^{2}}
+ \alpha Tr(\mathbf{Y}^T\mathbf{L}\mathbf{Y}),
\end{equation}}
where $\alpha$ is a balance parameter, which can be seen as a regularizer for the above  collective non-matrix factorization.

Intuitively, we learn coefficients of the optimal projection $\mathbf{W}^t$ by minimizing the error term between the linear hash function $ \mathbf{H}^t({\mathbf{X}}^t)=sgn({\mathbf{W}}^{t^T}{\mathbf{X}}^t)$ and the hash codes $\mathbf{Y}$ as $\left\| \mathbf{Y} - \mathbf{H}^t(\mathbf{X}^t) \right\|^2$.
Inspired by \cite{he2015optimal}, label graph can be used as Laplacian Regularized in this error.
Then, the objective function of hash function learning is written as follows:
\small{
\begin{equation}
\min_{\mathbf{H}^t(\mathbf{X}^t)} \left\| \mathbf{Y} - \mathbf{H}^t(\mathbf{X}^t) \right\|^2 + \eta Tr(\mathbf{H}^t(\mathbf{X}^t)\mathbf{L}\mathbf{H}^t(\mathbf{X}^t)^T),
\end{equation}}
where $\eta$ is the balance parameter.
This term is integrated with the term of supervised semantic similarity in the proposed graph-regularized collective matrix factorization. 
Then, the overall objective function is written as follows: 
\small{
\begin{eqnarray}\label{Eq7}
&\min_{\mathbf{Y},\mathbf{W}^t}\sum_{t=1}^{2} \lambda_t{\left \| {\mathbf{X}^t}-\mathbf{U}^t\mathbf{Y} \right \|_{F}^{2}}
+ \alpha Tr(\mathbf{Y}\mathbf{L}\mathbf{Y}^T)  \nonumber\\
&+ \mu \sum_{t=1}^{2} \big\{ \left \| \mathbf{Y} - \mathbf{H}^t({\mathbf{X}}^t) \right \|_{F}^{2} + \eta Tr(\mathbf{H}^t(\mathbf{X}^t)\mathbf{L}\mathbf{H}^t(\mathbf{X}^t)^T) \big\}\\
&s.t. \: \mathbf{Y} \in{\{0,1\}}^{r\times{N}}, 
	\: \mathbf{Y}\mathbf{Y}^T=\mathbf{I},
   \: \mathbf{U}^t \geq  0 ,\nonumber
\end{eqnarray}}
where $\mu$ is a tradeoff parameter to control the weights between hash function approximation and the proposed graph-regularized collective non-matrix factorization scheme.

\subsection{Optimization}
Directly minimizing the objective function in Eq. (\ref{Eq7}) is intractable due to the discrete constraint of $\mathbf{Y}$.
To tackle this issue, we relax the discrete constraint from $\mathbf{Y} \in{\{0,1\}}^{r\times{n}}$ to $0 \leq \mathbf{Y} \leq 1$. 
After that, it is still non-convex with respect to $\mathbf{Y},  \mathbf{U}^t$,  and $\mathbf{W}^t$ jointly.
This is further handled by using an alternating optimization, \textit{i.e.}, updating one variable while fixing the rest two until convergence.
Due to the scale of the label similarity matrix, it is inefficient to store and compute it during optimization.
To solve this problem, we propose a random sampling method, which uses a sampled sub-graph to replace the complete similarity matrix.
The detailed optimization procedure is presented as follows:

(1) \textbf{Fix $\mathbf{W}^t$ and $\mathbf{Y}$, then update $\mathbf{U}^t$. } The corresponding sub-problem is 
\small{
\begin{eqnarray}\label{Eq8}
&\min_{\mathbf{U}^1,\mathbf{U}^2}\sum_{t=1}^{2} \lambda_t{\left \| {\mathbf{X}}^t-\mathbf{U}^t\mathbf{Y} \right \|_{F}^{2}}\\
&s.t. \: \mathbf{U}^1 \geq 0,
   \: \mathbf{U}^2 \geq 0, \nonumber
   \: \lambda_1 + \lambda_2 = 1 .\nonumber
\end{eqnarray}}
Here we learn two basis matrices $\mathbf{U}^1$ and  $\mathbf{U}^2$, which turns this sub-problem to be a traditional NMF problem for each modality. 

By directly optimizing Eq. (\ref{Eq8}) within each modality respectively, we solve the new objective function through Lagrange multiplier $\Psi=[\psi_{ik}]$ with constant $u^t_{ij} \geq 0$:
\small{
\begin{equation}\label{Eq9}
\begin{split}
O_1(\mathbf{U}^t) & = Tr({\mathbf{X}}^t{{\mathbf{X}}}^{t^T}) - 2Tr({\mathbf{X}}^t\mathbf{Y}^T{\mathbf{U}^t}^T) \\
& + Tr(\mathbf{U}^t\mathbf{Y}\mathbf{Y}^T{\mathbf{U}}^{t^T}) + Tr(\Psi{\mathbf{U}}^{t^T}).
\end{split}
\end{equation}}
We derive the partial derivatives with respect to $\mathbf{U}^t$ as :
\small{
\begin{eqnarray}\label{Eq10}
\frac{\partial{O_1}}{\partial{\mathbf{U}^t}}
= -2\mathbf{X}^t\mathbf{Y}^T + 2\mathbf{U}^t\mathbf{Y}\mathbf{Y}^T + \Psi.
\end{eqnarray}}
Then by Karush-Kuhn-Tucker (KKT) conditions, we update $\mathbf{U}^t$ via the following procedure:
\small{
\begin{equation}\label{Eq11}
u^t_{ij} \leftarrow u^t_{ij}\frac{({\mathbf{X}}^t\mathbf{Y}^T)_{ij}}{(\mathbf{U}^t\mathbf{Y}\mathbf{Y}^T)_{ij}}.
\end{equation}} 
Intuitively, $\mathbf{\mathbf{U}}^t$ is non-negative after updating.

(2) \textbf{Fix $\mathbf{U}^t$ and $\mathbf{W}^t$, then update $\mathbf{Y}$. } We then fix $\mathbf{U}^t$ and $\mathbf{W}^t$, the updating of $\mathbf{Y}$ subsequently refers to:
\small{
\begin{eqnarray}\label{Eq12}
&\min_{\mathbf{Y}} \sum_{t=1}^{2} \lambda_t{\left \| {\mathbf{X}}^t-\mathbf{U}^t\mathbf{Y} \right \|_{F}^{2}} + \alpha Tr(\mathbf{Y}^T\mathbf{L}\mathbf{Y})  \nonumber \\
& + \mu \sum_{t=1}^{2}\left \| \mathbf{Y} - {\mathbf{W}^t}^T{\mathbf{X}}^t \right \|_{F}^{2} + \beta \| \mathbf{YY}^T-\mathbf{I}\|_{F}^{2} \\
&s.t.\quad0 \leq \mathbf{Y} \leq 1.\nonumber
\end{eqnarray}}
However, the scale of the semantic matrix is extremely large, which needs huge storage cost and makes the Eq. (\ref{Eq11}) hard to optimize during each iteration.
To tackle this problem, we randomly sample parts of the original label similarity matrix $A$, which can approximate the graph regularization.
We define a sampling matrix ${\mathbf{S}}\in{\{0,1\}^{N\times{m}}}$ during each iteration, where m is the number of sampling points with $m << N$.
Then the hash code of sampled data can be represented by $\hat{\mathbf{Y}}=\mathbf{Y}^T\mathbf{S}\in{\{0,1\}^{d_t\times{m}}}$, 
and the sampled label representation can be  presented by $\hat{\mathbf{L_a}}=\mathbf{L_a}^T\mathbf{S}\in{\{0,1\}^{dl\times{m}}}$.
By using the above sampling, this sub-problem of Eq. (\ref{Eq11}) can be rewritten as:
\small{
\begin{eqnarray}\label{Eq13}
&\min_{\mathbf{Y}} \sum_{t=1}^{2} \lambda_t{\left \| {\mathbf{X}}^t-\mathbf{U}^t\mathbf{Y} \right \|_{F}^{2}} + \alpha Tr(\hat{\mathbf{Y}}^T\hat{\mathbf{L}}\hat{\mathbf{Y}})  \nonumber \\
& + \mu \sum_{t=1}^{2}\left \| \mathbf{Y} - {\mathbf{W}^t}^T{\mathbf{X}}^t \right \|_{F}^{2} + \beta \| \mathbf{YY}^T-\mathbf{I}\|_{F}^{2} \\
&s.t. \quad 0 \leq \mathbf{Y} \leq 1.\nonumber
\end{eqnarray}}
where $\hat{\mathbf{L}}$ is the Laplacian matrix for $\hat{\mathbf{A}}=\hat{\mathbf{L_a}}^T{\cdot}\hat{\mathbf{L_a}}$.

Since solving such a constraint is not convenient, we further relax it to $\mathbf{Y} \geq 0$ and normalize $\mathbf{Y}$ after factorization in each iteration.
Let $\Phi$ be the Lagrange Multiplier for the new constraint. 
The Lagrange term could be written as follows:
\small{
\begin{equation}\label{Eq14}
\begin{split}
 O_2(\mathbf{Y}) &= 
\sum_{t=1}^{2}\lambda_t\{Tr({\mathbf{X}}^t{\mathbf{X}}^{t^T}) - 2Tr({\mathbf{X}}^t\mathbf{Y}^T{\mathbf{U}}^{t^T}) \\
& + Tr(\mathbf{U}^t\mathbf{Y}\mathbf{Y}^T{\mathbf{U}^t}^T)\} + \alpha Tr(\hat{\mathbf{Y}}\hat{\mathbf{L}}\hat{\mathbf{Y}}^T) \\
& + \mu\sum_{t=1}^{2}\{Tr(\mathbf{Y}\mathbf{Y}^T)-2Tr(\mathbf{Y}{\mathbf{X}}^{t^T}\mathbf{W}^t) \\
& + Tr({\mathbf{W}}^{t^T}{\mathbf{X}}^t{\mathbf{X}}^{t^T}\mathbf{W}^t)\}+ \beta \{ Tr(\mathbf{Y}\mathbf{Y}^T\mathbf{Y}\mathbf{Y}^T) \\
& - 2Tr(\mathbf{Y}\mathbf{Y}^T) + const \} + Tr(\Phi \mathbf{Y}^T). \\
\end{split}
\end{equation}}
Using the KKT conditions, we have:
\small{
\begin{equation}\label{Eq15}
\begin{split}
&\begin{split}
\frac{\partial{O_2}}{\partial{\mathbf{Y}}}
&= 2\sum_{t=1}^{2}{\lambda_t}{\{-\mathbf{U}}^{t^T}{\mathbf{X}}^t + {\mathbf{U}}^{t^T}\mathbf{U}^t\mathbf{Y} \}+ 2\alpha{\hat{\mathbf{Y}}\hat{\mathbf{L}}\mathbf{S}^T}\\
& + 2\mu\sum_{t=1}^{2}\{\mathbf{Y} - {\mathbf{W}}^{t^T}{\mathbf{X}}^t\} + 4\beta\{\mathbf{Y}\mathbf{Y}^T\mathbf{Y}-\mathbf{Y} \}+ \Phi = 0. \\
& s.t. \ \psi_{ij} \mathbf{Y}_{ij}=0, \mathbf{Y} \geq 0.
\end{split}
\end{split}
\end{equation}}
Eq. (\ref{Eq13}) can be solved by the following updating rule:
\small{
\begin{equation}\label{Eq16}
\begin{split}
& y_{ij} \leftarrow \\
& y_{ij}\frac{(\sum_{t=1}^{2}{\lambda_t}{\mathbf{U}^t}^T{\mathbf{X}}^t+\alpha \hat{\mathbf{Y}}\hat{\mathbf{A}}\mathbf{S}^T+\mu\sum_{t=1}^{2}{\mathbf{W}}^{t^T}{\mathbf{X}}^t+4\beta\mathbf{Y})_{ij}}{(\sum_{t=1}^{2}{\lambda_t}{{\mathbf{U}^t}^T\mathbf{U}^t\mathbf{Y}}+\alpha \hat{\mathbf{Y}}\hat{\mathbf{D}}\mathbf{S}^T + \mu\sum_{t=1}^{2}{\mathbf{Y}}+4\beta\mathbf{Y}(\mathbf{Y}^T\mathbf{Y}))_{ij}}.
\end{split}
\end{equation}}
$\hat{\mathbf{D}}$ is the diagonal matrix with entries of column sums of $\hat{\mathbf{A}}$.

(3) \textbf{Fix $\mathbf{U}^t$ and $\mathbf{Y}$, then update $\mathbf{W}^t$. } This last sub-problem finds the best projection coefficient $\mathbf{W}^t$ by minimizing Eq. (7) for the $t$-th modality as the Laplacian Regularized Least squares algorithm \cite{he2015optimal}, 
resulting in a closed-form solution:
%
%
%
\small{
\begin{equation}\label{Eq18}
\mathbf{W}^t = \big({\mathbf{X}}^t{\mathbf{X}}^{t^T} + \eta ({\mathbf{X}}^t \mathbf{S})\mathbf{\hat{L}}({\mathbf{S}^T}{\mathbf{X}}^{t^T}) +\gamma \mathbf{I}\big)^{-1}{\mathbf{X}}^t\mathbf{Y}^T.
\end{equation}}

\begin{algorithm}[t]
\caption{ Supervised Matrix Factorization Hashing}
\renewcommand{\algorithmicrequire}{\textbf{Input:}} 
\renewcommand{\algorithmicensure}{\textbf{Output:}}
\begin{algorithmic}[1]
\REQUIRE 
Training data points in two modalities $\mathbf{X}^1$ and $\mathbf{X}^2$,
the corresponding pairwise semantic similarity matrix $\mathbf{A}$,
the number of support samples \textit{m}, and the number of hash bits \textit{r}.
\ENSURE 
The hash codes $\mathbf{Y}$ for training data and the projection coefficient  matrix $\mathbf{W}^t$.\\
\STATE
Initialize $\mathbf{W}^t$, $\mathbf{U}^t$ and $\mathbf{Y}$ by random matrices, $t=1,2$.
\REPEAT 
\STATE 
 Fixing $\mathbf{W}^t$ and $\mathbf{Y}$, update $\mathbf{U}^t$ by Eq.  (\ref{Eq11});\\ 
 \STATE
Uniformly and randomly select \textit{m} sample pairs from training data. \\
\STATE
 Fixing $\mathbf{U}^t$ and $\mathbf{W}^t$, update $\mathbf{Y}$ by Eq.  (\ref{Eq16});\\ 
\STATE
 Fixing $\mathbf{U}^t$ and $\mathbf{Y}$, update $\mathbf{W}^t$ by Eq.  (\ref{Eq18});\\
\UNTIL{convergence} 
\end{algorithmic}
\label{alg1}
\end{algorithm}

We summarize the whole procedure of the proposed SMFH in Algorithm \ref{alg1}.

%
\textbf{Times Complexity:} The main time consumption of the proposed SMFH is the matrix factorization, its complexity is $O\Big( \big( ndr+(n+d)(r^2+r)+m^2r\big)t \Big)$, where $t$ is the number of iterations.
Since $r,m << n$, the overall complexity is $O\big(n(d+r)rt\big)$, which is linear to the size of the training data.
In practice, the proposed SMFH is much faster than most cross-modality hashing, \textit{i.e.} PDH and CMFH, whose training time is rounded to the competing cross-moality hashing method SMH.
Although the hash codes of training data is obtained by minimizing Eq. (\ref{Eq7}), it cannot be directly applied to the case of the out-of-sample query.
For such out-of-sample query, we use the hash function learned by Eq. (\ref{Eq7}) to generate the corresponding binary code.
Then in online search, the time complexity for each modality is constant as $O(dr)$.

\begin{table*}[!tbh]
\centering
\caption{The \emph{m}AP and Precision Comparison Using Hamming Ranking on Two Benchmark with Different Hash Bits.}
\label{my-label}
\scalebox{0.95}[0.95]{
\begin{tabular}{|c|c|c|c|c|c|c|c|c|c|c|c|}
\hline
\multirow{3}{*}{Task}            & \multirow{3}{*}{Methods} & \multicolumn{5}{c|}{Wiki}                                                                & \multicolumn{5}{c|}{NUS-WIDE}                                                            \\ \cline{3-12} 
                                 &                          & \multicolumn{3}{c|}{\emph{m}AP}                            & \multicolumn{2}{c|}{Precision@100} & \multicolumn{3}{c|}{\emph{m}AP}                            & \multicolumn{2}{c|}{Precision@100} \\ \cline{3-12} 
                                 &                          & 32              & 64              & 128             & 32               & 64              & 32              & 64              & 128             & 32               & 64              \\ \hline
\multirow{5}{*}{\textbf{Task 1}} & CVH                      & 0.2053          & 0.1872          & 0.2039          & 0.1504           & 0.1324          & 0.4480          & 0.4184          & 0.4012          & 0.4606           & 0.4281          \\ \cline{2-12} 
                                 & PDH                      & 0.2034          & 0.2047          & 0.2133          & 0.1765           & 0.1725          & 0.5008          & 0.5078          & 0.5336          & 0.4885           & 0.5018          \\ \cline{2-12} 
                                 & CMFH                     & 0.5947          & 0.6063          & 0.6131          & 0.5492           & 0.5649          & 0.3807          & 0.3787          & 0.3663          & 0.3764           & 0.3750          \\ \cline{2-12} 
                                 & SMH                      & 0.3831          & 0.4032          & 0.4171          & 0.3135           & 0.3400          & \textbf{0.5978}          & 0.6162          & 0.6195          & \textbf{0.6007}           & 0.6194          \\ \cline{2-12} 
                                 & \textbf{SMFH}            & \textbf{0.6039} & \textbf{0.6602} & \textbf{0.6658} & \textbf{0.5581}  & \textbf{0.6246} & {0.5462} & \textbf{0.6633} & \textbf{0.6247} & {0.5478}  & \textbf{0.6757} \\ \hline
\multirow{5}{*}{\textbf{Task 2}} & CVH                      & 0.1660          & 0.1479          & 0.1572          & 0.1280           & 0.1170          & 0.4592          & 0.4260          & 0.4021          & 0.4700           & 0.4333          \\ \cline{2-12} 
                                 & PDH                      & 0.2442          & 0.2360          & \textbf{0.2685}          & 0.2058           & 0.1945          & 0.5129          & 0.5260          & 0.5377          & 0.4988           & 0.5224          \\ \cline{2-12} 
                                 & CMFH                     & 0.2081          & 0.2111          & 0.2270          & 0.1691           & 0.1678          & 0.3818          & 0.3774          & 0.3664          & 0.3787           & 0.3766          \\ \cline{2-12} 
                                 & SMH                      & 0.2301          & 0.2503          & 0.2570          & 0.1953           & 0.2186          & {0.5823}          & {0.6020}          & {0.6089}          & {0.5829}           & {0.6039}          \\ \cline{2-12} 
                                 & \textbf{SMFH}            & \textbf{0.2516} & \textbf{0.2581} & {0.2496} & \textbf{0.2168}  & \textbf{0.2330}          & \textbf{0.5938} & \textbf{0.6325} & \textbf{0.6175} & \textbf{0.5985}  & \textbf{0.6380} \\ \hline
\end{tabular}}
\end{table*}

\subsection{Extension to Multi-Modality Search}

It is quite intuitive to extend SMFH in Eq. (\ref{Eq7}) from bi-modal to multiple modalities, that is:
\small{
\begin{eqnarray}\label{Eq19}
&\min_{\mathbf{Y},\mathbf{W}^t}\sum_{t=1}^{n_t} \lambda_t{\left \| {\mathbf{X}^t}-\mathbf{U}^t\mathbf{Y} \right \|_{F}^{2}}
+ \alpha Tr(\mathbf{Y}\mathbf{L}\mathbf{Y}^T)  \\
&+ \mu \sum_{t=1}^{n_t} \{ \left \| \mathbf{Y} - \mathbf{H}^t({\mathbf{X}}^t) \right \|_{F}^{2} + \eta Tr(\mathbf{H}^t(\mathbf{X}^t)\mathbf{L}\mathbf{H}^t(\mathbf{X}^t)^T) \} \nonumber\\
&s.t. \: \mathbf{Y} \in{\{0,1\}}^{r\times{n}}, 
   \: \mathbf{U}^t \geq  0 ,\nonumber
\end{eqnarray}}
where $\sum_{t=1}^{n_t}\lambda_t=1$.
It is convenient to adopt Algorithm \ref{alg1} to minimize the objective function in Eq. (\ref{Eq7}).
An alternating optimization  strategy can also be used here.
In particular, the variable $\mathbf{U}^t$ and $\mathbf{W}^t$ can be directly  got through Eq. (\ref{Eq11}) and Eq. (\ref{Eq16}), respectively.%
And finally, the variable $\mathbf{Y}$ can be learned by the new formulation as follows:
\small{
\begin{equation}\label{Eq20}
\begin{split}
& y_{ij} \leftarrow \\
& y_{ij}\frac{(\sum_{t=1}^{n_t}{\lambda_t}{\mathbf{U}^t}^T{\mathbf{X}}^t+\alpha \hat{\mathbf{Y}}\hat{\mathbf{A}}\mathbf{S}^T+\mu\sum_{t=1}^{n_t}{\mathbf{W}}^{t^T}{\mathbf{X}}^t++4\beta\mathbf{Y})_{ij}}{(\sum_{t=1}^{n_t}{\lambda_t}{{\mathbf{U}^t}^T\mathbf{U}^t\mathbf{Y}}+\alpha \hat{\mathbf{Y}}\hat{\mathbf{D}}\mathbf{S}^T + \mu\sum_{t=1}^{n_t}{\mathbf{Y}} +4\beta\mathbf{Y}(\mathbf{Y}^T\mathbf{Y}))_{ij}}.
\end{split}
\end{equation}}


\section{Experiments}\label{Sec4}

Quantitative experiments are conducted to validate the advantages of the proposed cross-modality hashing algorithm on three widely-used  benchmark, \emph{i.e.}, \textbf{PASCAL-Sentence}\footnote{http://vision.cs.uiuc.edu/pascal-sentences/}, \textbf{Wiki}\footnote{http : //www.svcl.ucsd.edu/projects/crossmodal/} and \textbf{NUS-WIDE}\footnote{http : //lms.comp.nus.edu.sg/research/NUS−WIDE.htm}.

The \textbf{PASCAL-Sentence} dataset contains 1,000 images that are divided into 20 categories. 
Each image is represented by a 269-dimensional visual feature extracted by a collections detectors.
A 2,790-dimensional textual feature is extracted using the bag-of-words representation with WordNet \cite{farhadi2010every}.
For this dataset, 800 image-sentence pairs are randomly sampled as the training set and the remaining for query testing.

The \textbf{Wiki} dataset contains 2,866 documents, where the image-text pairs are fully annotated with 10 semantic categories.
Each image is represented as a 128-dimensional bag-of-visual-words feature. 
Each document is represented as a 10-dimensional topical feature   using Latent Dirichlet Allocation \cite{blei2003latent}.
For the Wiki dataset, we randomly select $75\%$ image-text pairs for training and the rest for query testing.

The \textbf{NUS-WIDE} dataset contains 269,648 images with 81 concepts crawled from Flickr.
We select 186,577 labeled image-text pairs according to the top 10 largest concepts as adopted in \cite{hu2014iterative,zhang2014large}.
In this dataset, images are represented by a 500-dimensional bag-of-visual-words feature, and its corresponding tags are represented by a 1,000-dimensional bag-of-words feature.
We choose 6,577 image-text pairs from this database as query and the remaining to form the dataset for training.

\textbf{Compared Methods:} We evaluate the cross-modality retrieval task via: (1) the text-to-image side, termed \textbf{Task 1}, and (2) the image-to-text side, termed \textbf{Task 2}.
In both sides, the proposed \textbf{SMFH} is compared against four state-of-the-art methods: Cross-View Hashing (\textbf{CVH}) \cite{kumar2011learning},  Supervised MultiModal Hashing (\textbf{SMH}) \cite{zhang2014large}, Predictable Dual-view Hashing (\textbf{PDH})  \cite{rastegari2013predictable} and Collective Matrix Factorization Hashing (\textbf{CMFH})  \cite{ding2014collective}.
Except CVH, the source codes of the rest methods are available publicly, and all their parameters' setting are used as what their papers presented.
All our experiments were run on a workstation with a 3.60GHz Intel Core I5-4790 CPU and 16GB RAM.

\begin{figure}[t]
\begin{center}
\begin{minipage}[t]{0.6\linewidth}
\centerline{
\subfigure[The \emph{m}AP on Task 1.]{
\includegraphics[width=\linewidth]{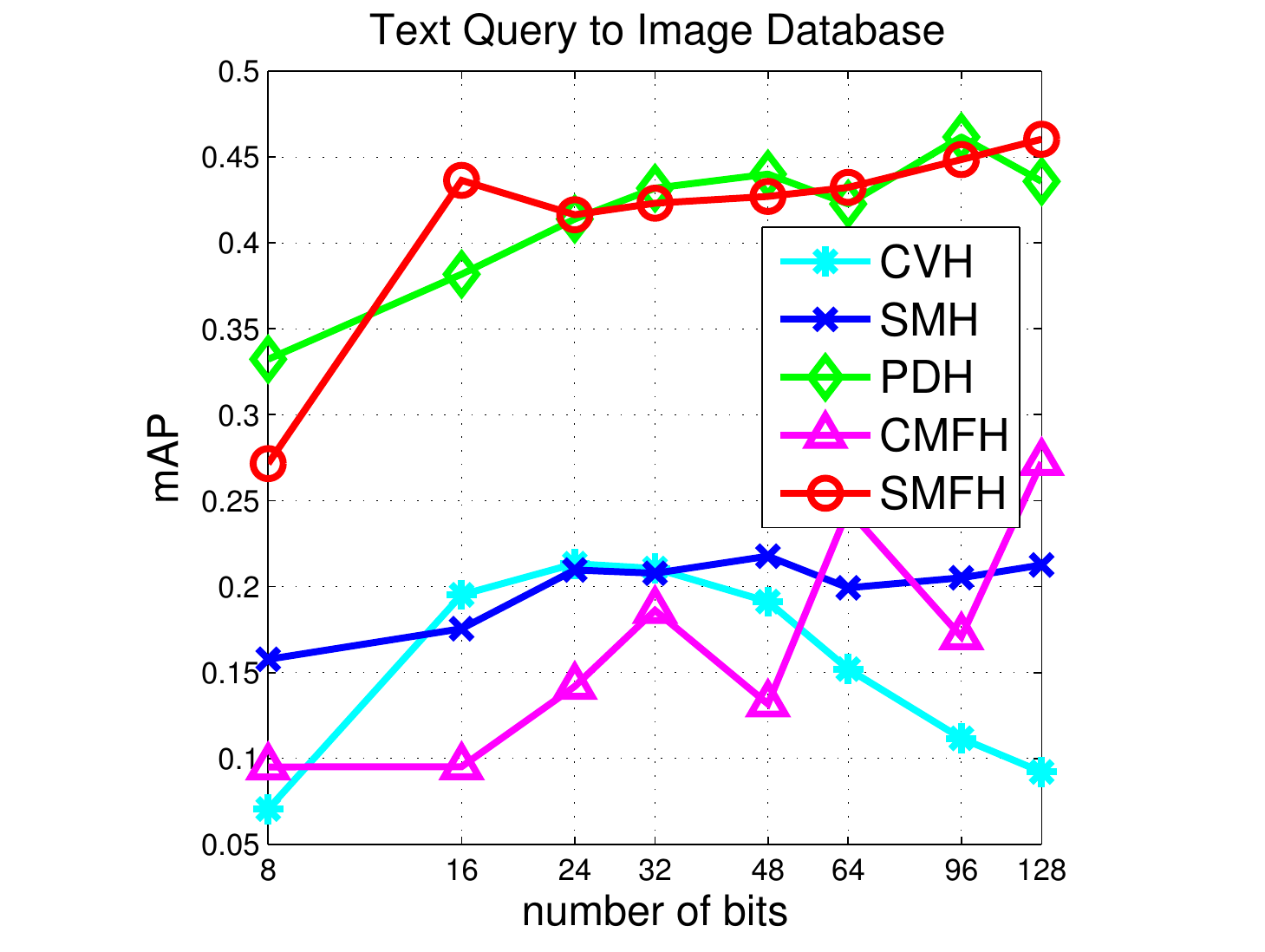}}\hspace*{-0.18\linewidth}
\subfigure[The \emph{m}AP on Task 2.]{
\includegraphics[width=\linewidth]{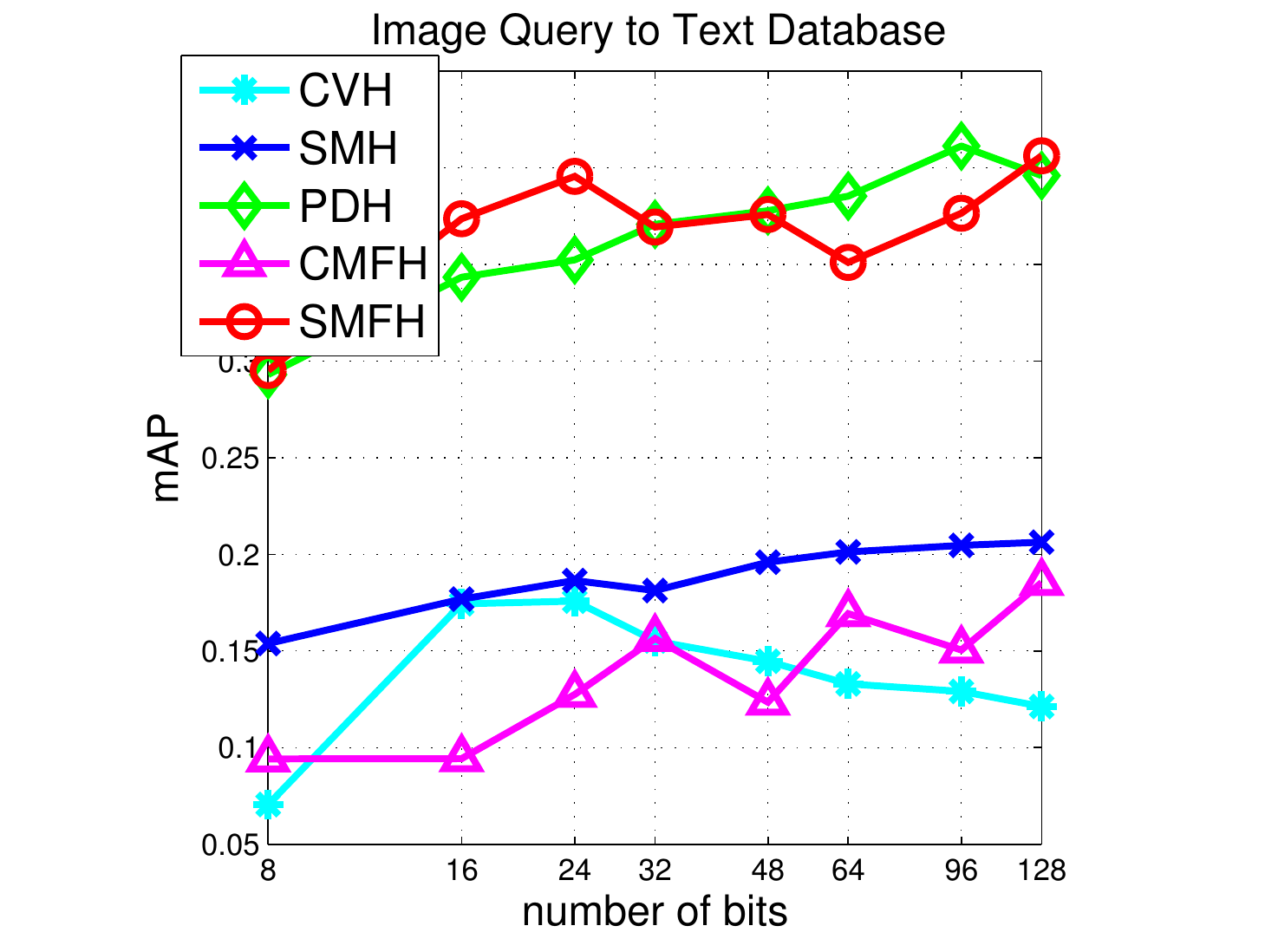}}
}
\end{minipage}
\end{center}
\setlength{\abovecaptionskip}{-5pt}
\caption{The \emph{m}AP curves on PASCAL-Sentence.}
\label{fig4}
\end{figure}

\textbf{Evaluation Protocols:} The quantitative performance is evaluated by using the mean Average Precision (\emph{m}AP) with top $100$ ranking list. 
We also consider other three evaluation protocols, \textit{i.e.}, precision at top-100 positions (Precision@100), Recall curves at top-K and Precision curves at top-K.

\textbf{Parameter Settings:} SMFH has five essential parameters in Eq. (\ref{Eq12}), \emph{i.e.},  $\lambda_1$, $\lambda_2$, $\alpha$, $\mu$, $\eta$ and $m$. 
The parameter $\lambda_1$ and $\lambda_2$ control the weights between two modalities, which are found to have little influence. 
In our experiments, we empirically set $\lambda_1=0.5$ for the image modality and $\lambda_2=0.5$ for the text modality.
The parameter $\alpha$ holds the semantic similarity of the original space, which is set as a large number of $2$.
$\mu$ is a trade-off parameter, which is set as $25$ on the two datasets.
During each iteration, the number of sampling points is set as $800$ for PASCAL-Sentence, $1,000$ for Wiki, and $2,000$ for the NUS-WIDE.
At the last part of this section, we will analyze the relation of the parameter $m$, and show the convergence result.
The three regularization parameters $\gamma$, $\beta$, and $\eta$ are set to a small number $0.001$ in all the experiments. 


\textbf{Quantitative Results:}
Fig. 2 shows the \emph{m}AP results on PASCAL-Sentence dataset with different bits on both retrieval tasks.
SMFH has achieved remarkable \emph{m}AP scores, especially when hash bit is larger than 32.
Comparing to the second best scheme, SMFH has achieved $7.1\%$ \emph{m}AP improvement for the Task 1 and $6.5\%$ improvement for the Task 2.

Then, we evaluate the proposed method on Wiki, as shown in the first row of Fig. 3 and Tab. 1, which demonstrate that SMFH has achieved superior performance on this benchmark for both text-to-image and image-to-text sides, both with a performance gain of more than $6\%$.
The \emph{m}AP results and Precision$@100$ results on Wiki are reported in Tab.1 under the setting of 32, 64, and 128 bits respectively.
SMFH has achieved remarkable \emph{m}AP and precision scores. 
Comparing with the state-of-the-art alogrithms, \emph{i.e.}, \cite{kumar2011learning,rastegari2013predictable,zhang2014large,ding2014collective}, 
for the task of text-to-image retrieval, our SMFH has significant advantage on precision and \emph{m}AP values with all bits, mainly due to the fact that the matrix factorization can successfully find better latent topic concepts from text. Meanwhile SMFH fully uses supervised label to  improve the cross-modality retrieval.
Fig.3 shows the comparison of precision curves and recall curves on Wiki when hash bit is 64.
\begin{figure*}[!htb]
\begin{center}

\begin{minipage}[t]{0.3\linewidth}
\centerline{
\subfigure[Precsion$@K$ on \textbf{Task 1}.]{
\includegraphics[width=\linewidth]{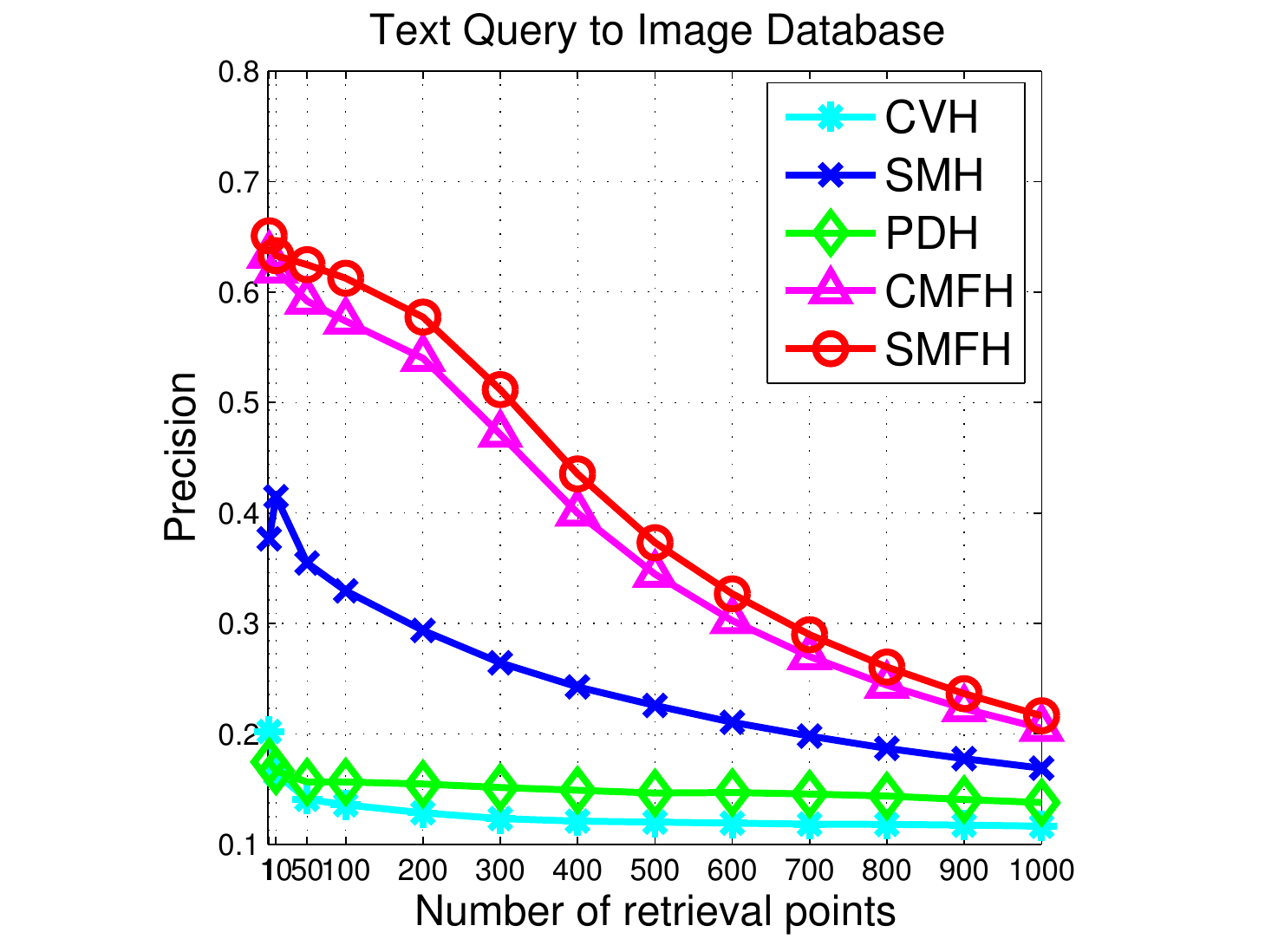}}\hspace*{-0.18\linewidth}
\subfigure[Recall$@K$ on \textbf{Task 1}.]{
\includegraphics[width=\linewidth]{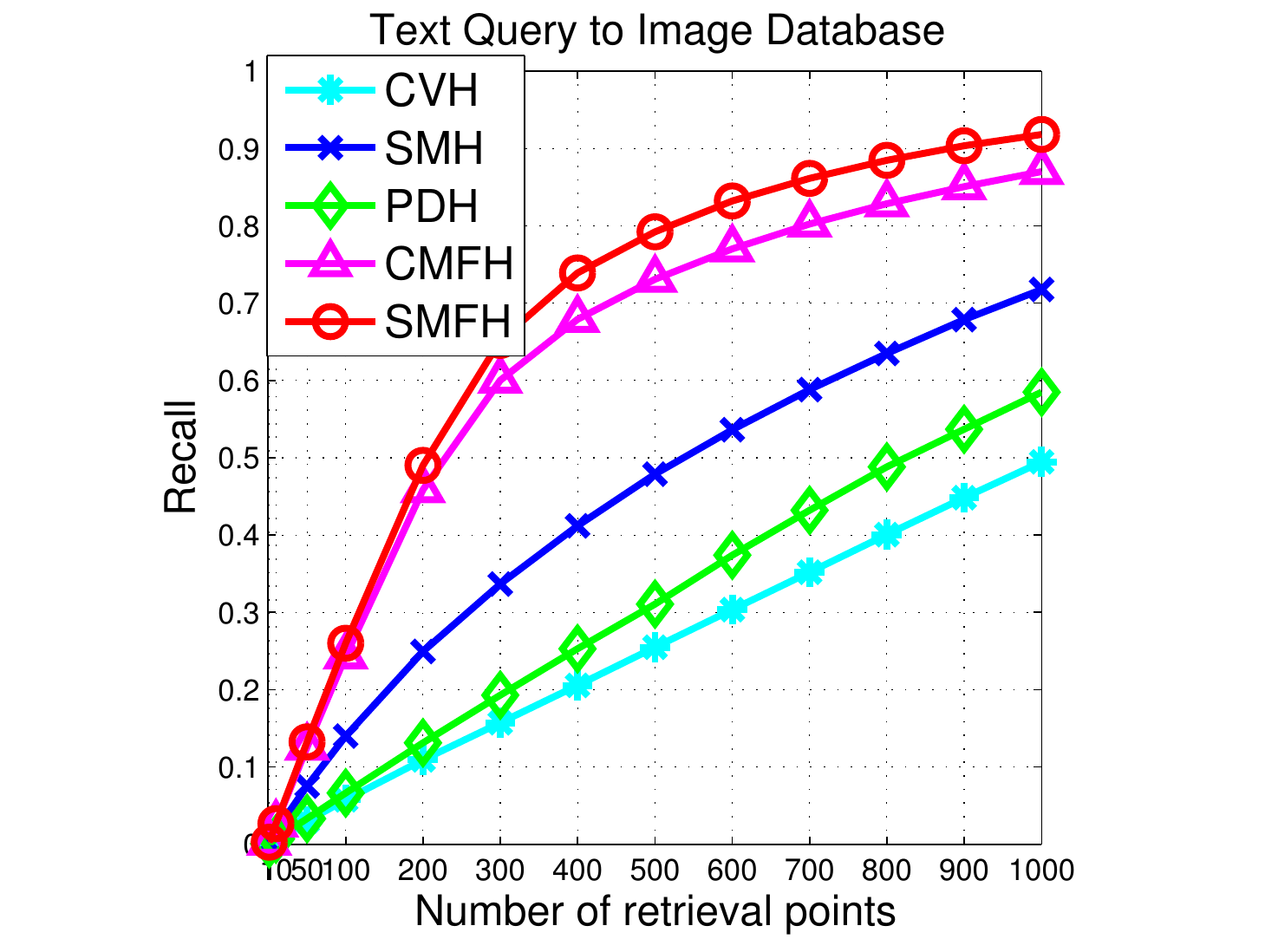}}\hspace*{-0.18\linewidth}
\subfigure[Precsion$@K$ on \textbf{Task 2}.]{
\includegraphics[width=\linewidth]{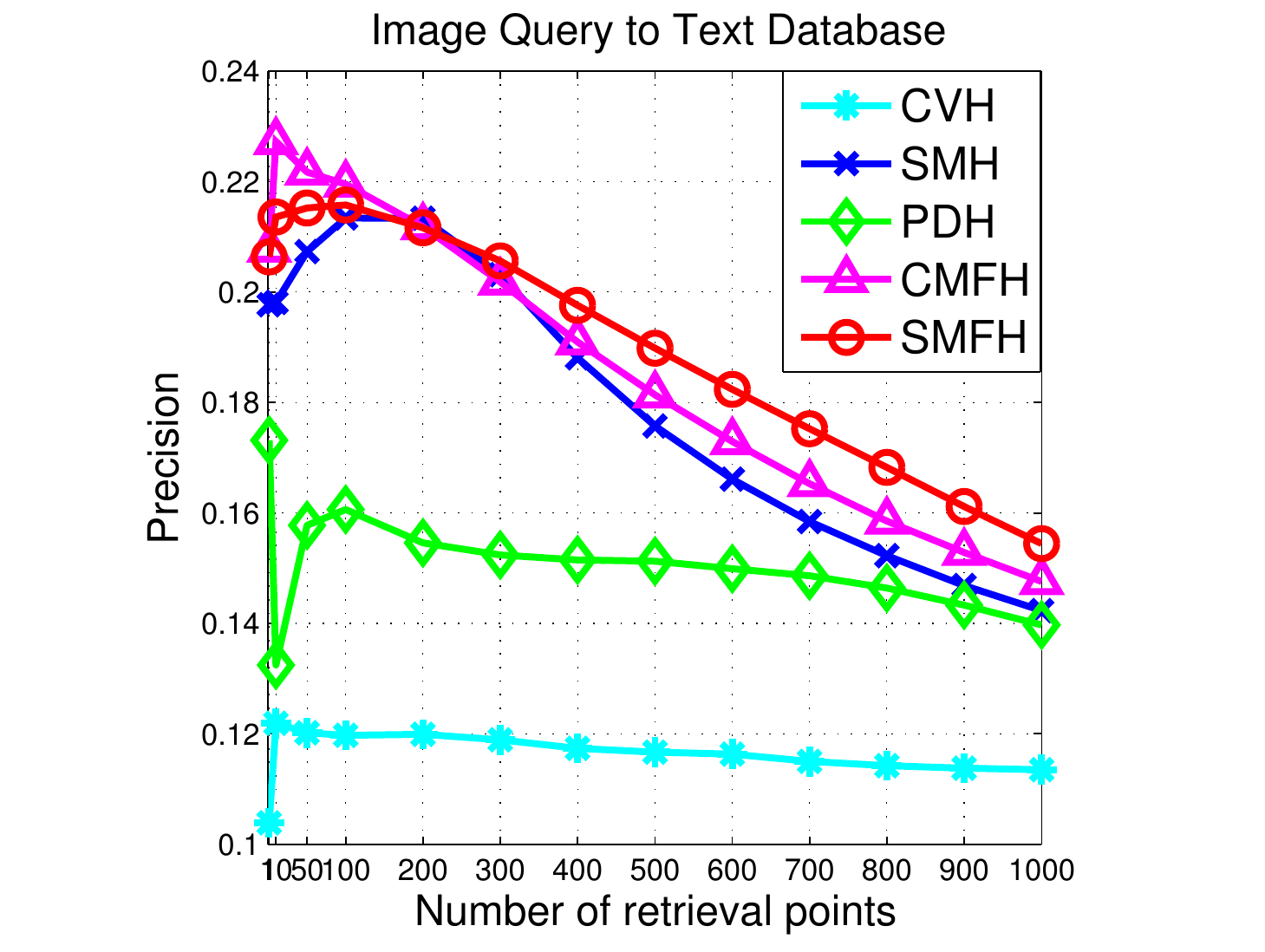}}\hspace*{-0.18\linewidth}
\subfigure[Recall$@K$ on \textbf{Task 2}.]{
\includegraphics[width=\linewidth]{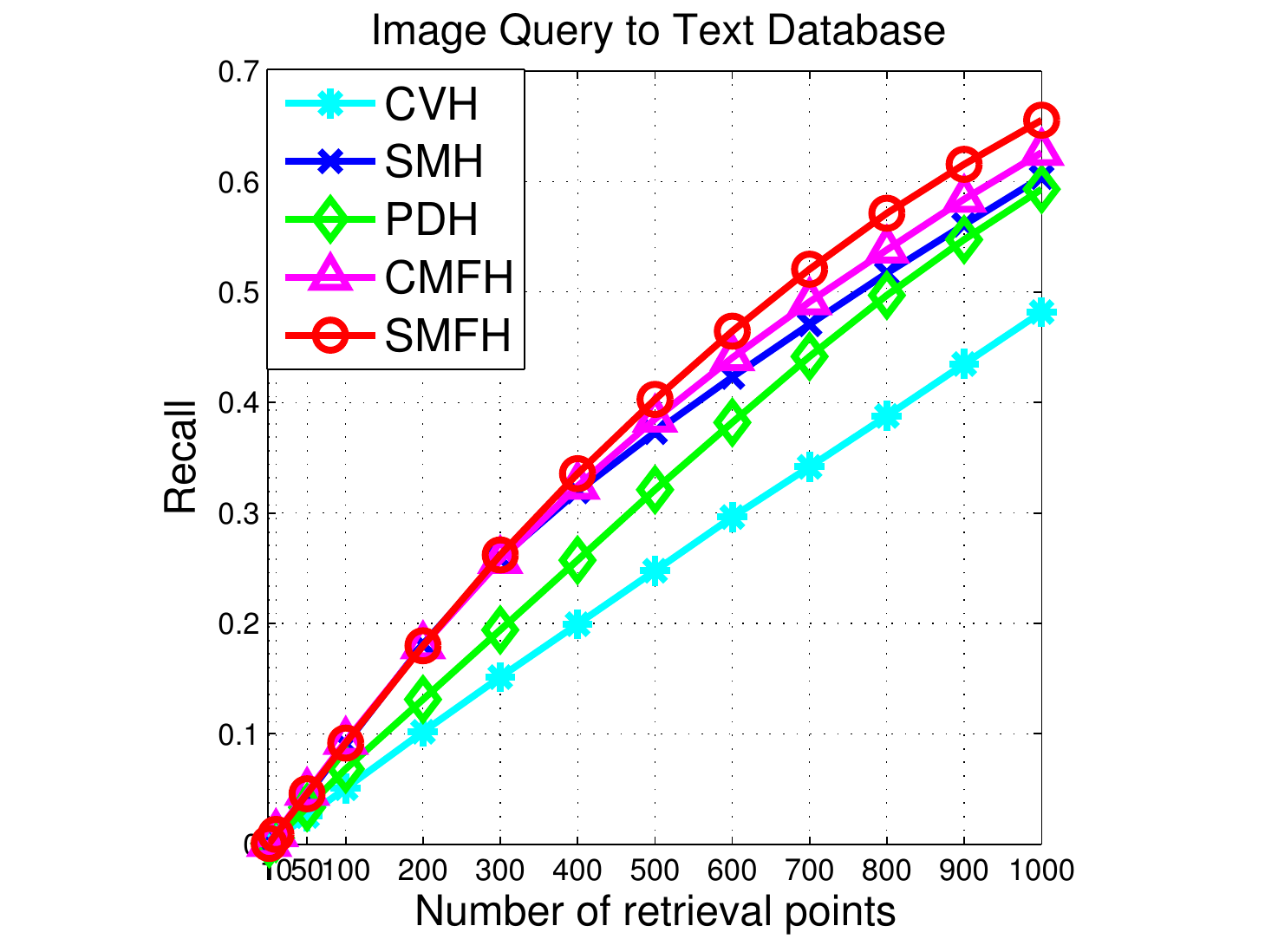}}
}
\centerline{\small (1) Wiki Dataset}
\end{minipage}

\begin{minipage}[t]{0.3\linewidth}
\centerline{
\subfigure[Precsion$@K$ on \textbf{Task 1}.]{
\includegraphics[width=\linewidth]{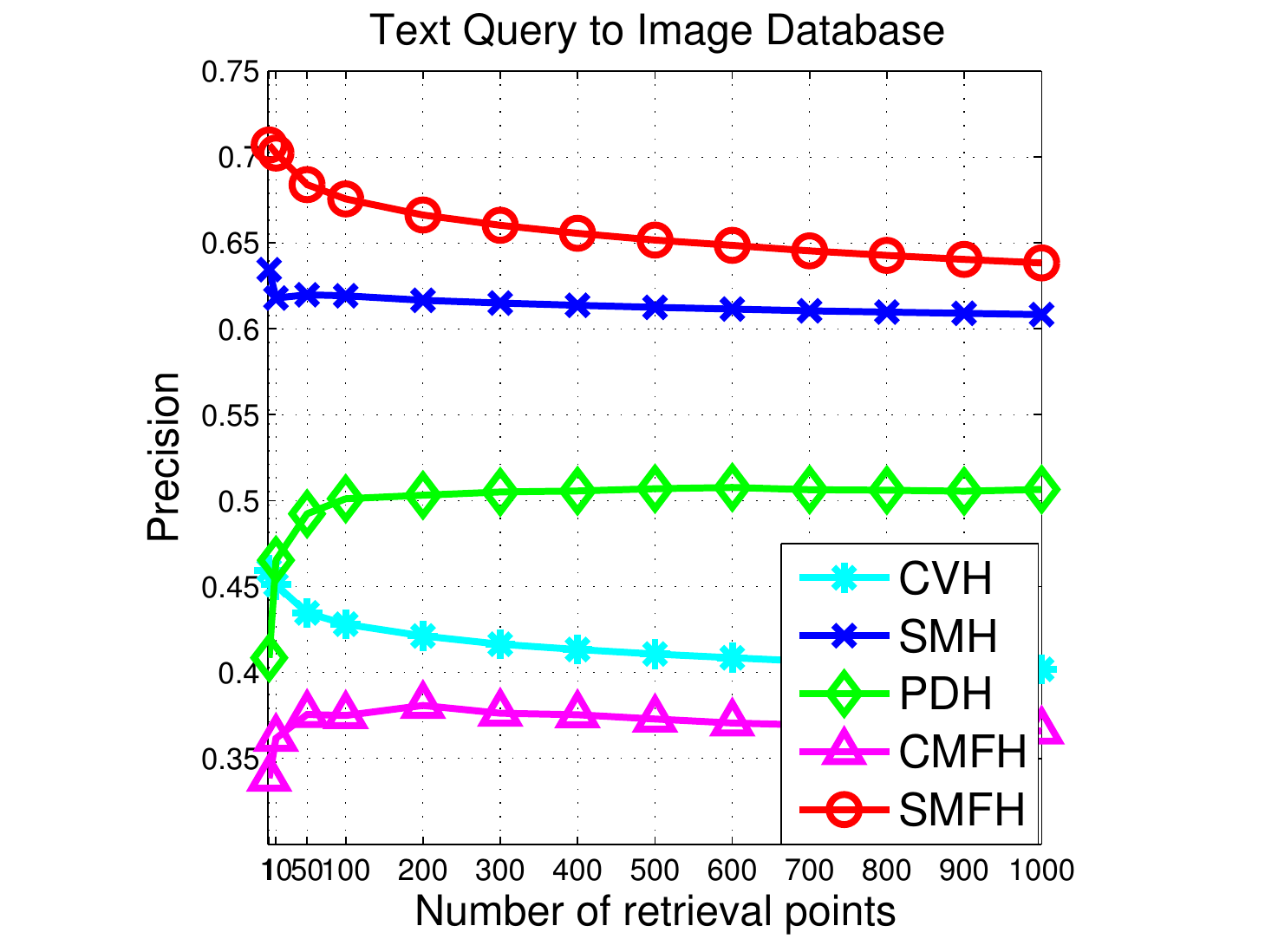}}\hspace*{-0.18\linewidth}
\subfigure[Recall$@K$ on \textbf{Task 1}.]{
\includegraphics[width=\linewidth]{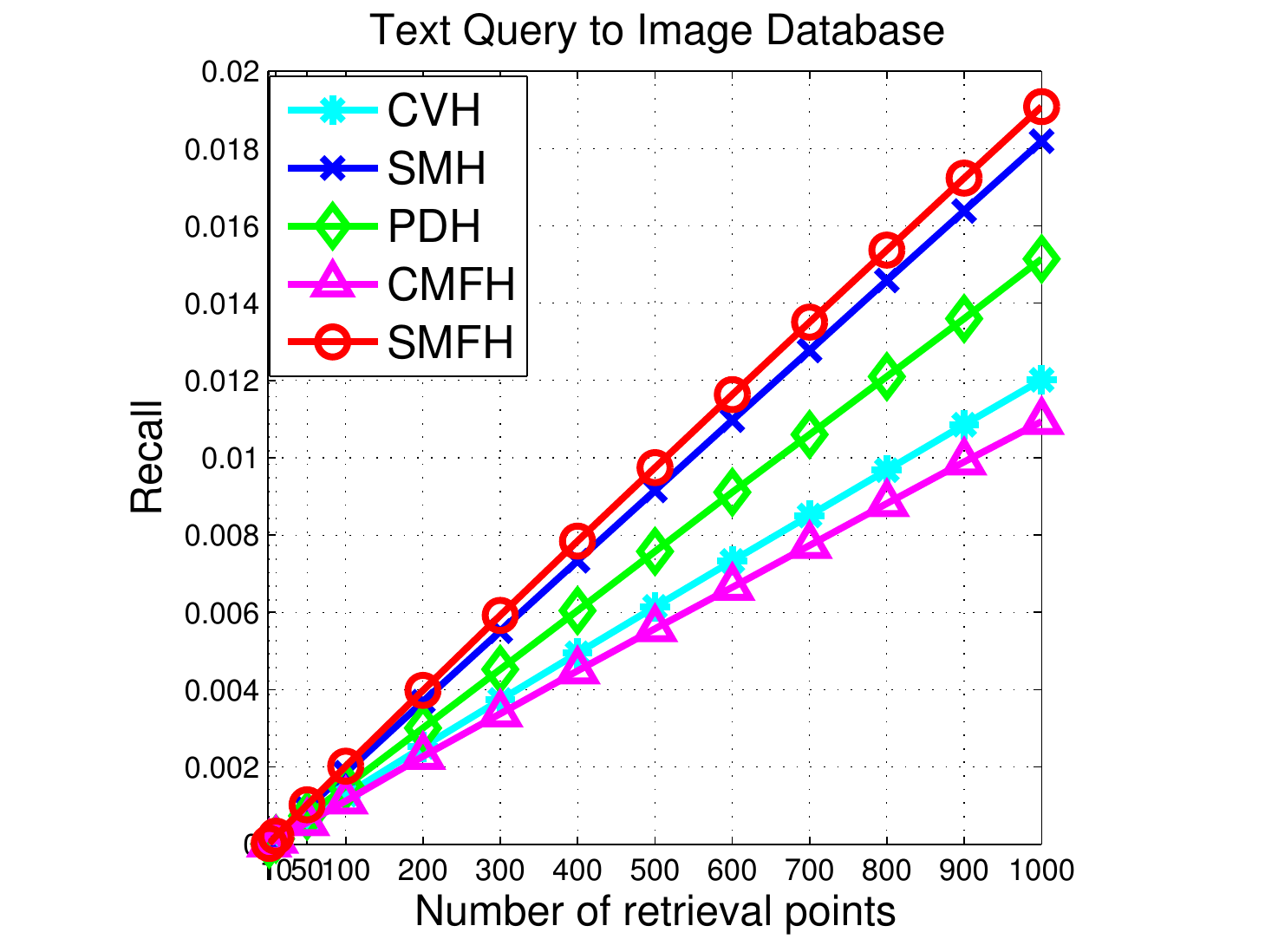}}\hspace*{-0.18\linewidth}
\subfigure[Precsion$@K$ on \textbf{Task 2}.]{
\includegraphics[width=\linewidth]{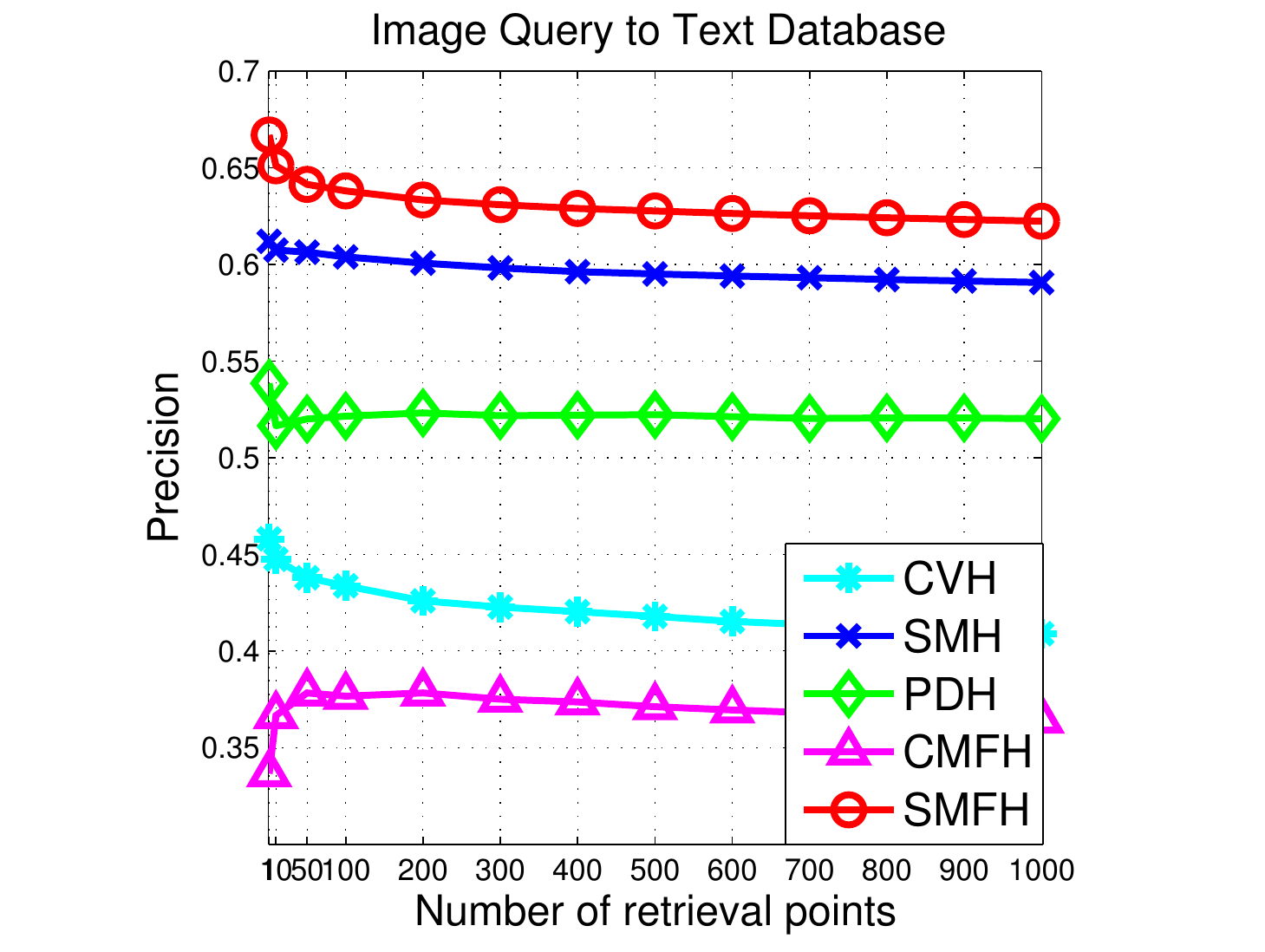}}\hspace*{-0.18\linewidth}
\subfigure[Recall$@K$ on \textbf{Task 2}.]{
\includegraphics[width=\linewidth]{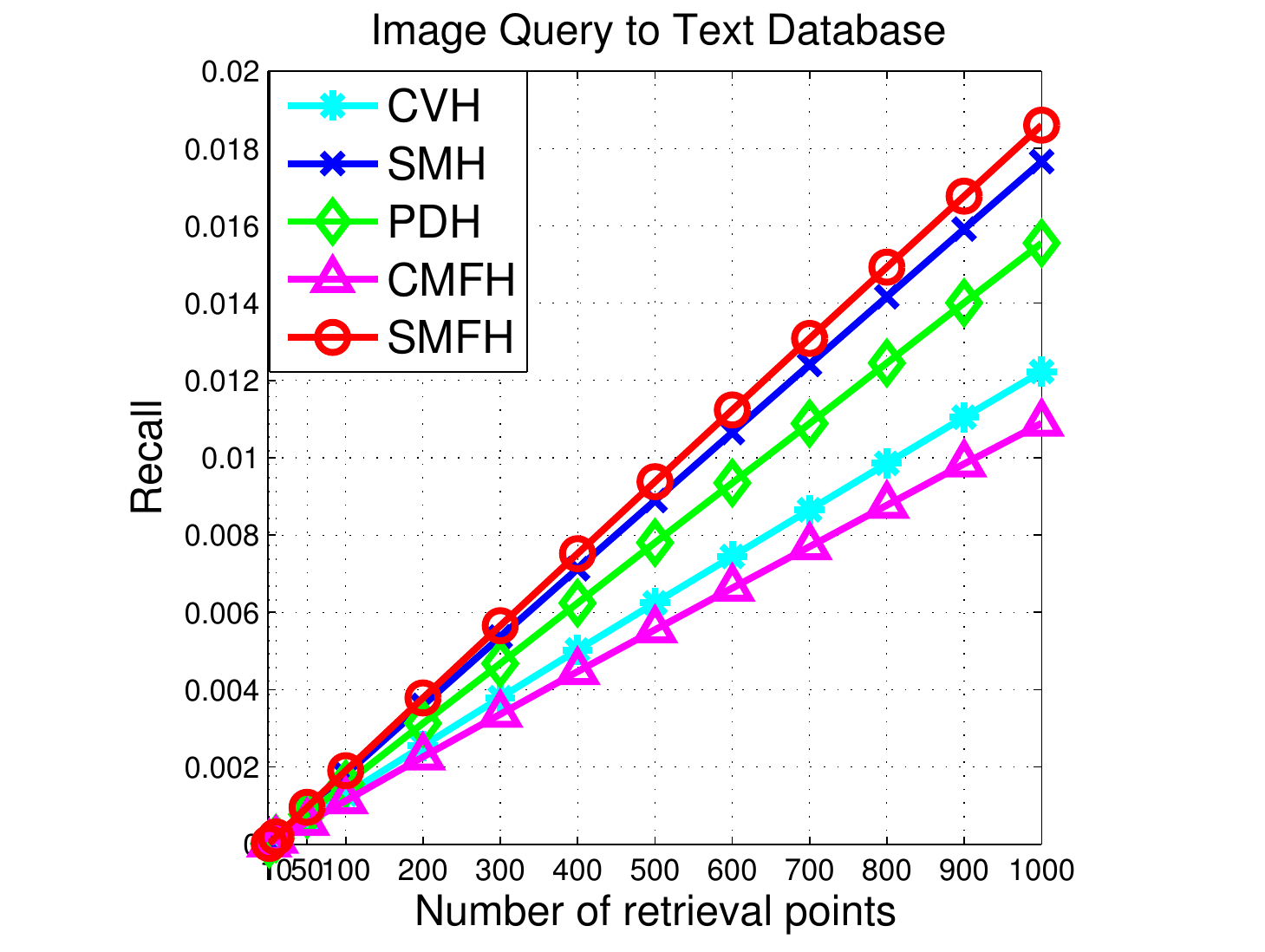}}
}
\centerline{\small (2) NUS-WIDE Dataset}
\end{minipage}
\end{center}
\setlength{\abovecaptionskip}{-2pt}
\caption{The Precision curves and Recall curves of all the algorithms on both datasets when hash bit is $64$.}
\label{fig2}
\end{figure*}

Similar performance gains are observed on the large-scale NUS-WIDE, especially in the text-to-image retrieval.
As shown in Tab.1 and the second row of Fig.3, SMFH achieves highest search accuracy. 
When the hash bit is 64, the precision of the top-100 Hamming ranking is over $60\%$ by SMFH, which is much better than the state-of-the-arts \cite{ding2014collective,kumar2011learning,rastegari2013predictable,zhang2014large}.
Although, the \emph{m}AP is not the highest when hash bit is $32$ for text-to-image sides, SMFH maintain its the competitive advantage on higher hash bits.
Nevertheless, the performance of Task 2 is at the second place, which also have competitive performance for cross-modality retrieval.

Tab.2 shows the results of training time comparing with different algorithms on different hash bits on three benchmarks, which contain the whole training set.
PDH and CMFH are much higher than that of our proposed SMFH, with larger size of training data and higher feature dimensions.
SMFH can get the better performance comparing with others, by fully use of the semantic information to enhance the performance with less training time. 

We further study the influence of different sizes of the training set. 
As shown in Fig.4 (a) for Wiki, \emph{m}AP results are shown when hash bit is 64, in which we vary the size of sampling from $200$ to $2,000$.
The performance of cross-modality retrieval consistently improve with the increasing of the sampling size for SMFH.
Thus, we randomly choose about 100 image-tag pairs for each concept as the training set for convenience in optimization, which contain about $1,000$ pairs during each iteration.
At last, we validate the convergence according to the sampling during each iteration.
As shown in Fig.4 (b), when the size of sampling pairs is $1,000$, and the hash bit is 64, SMFH can quickly  converge by using random sampling.
The same conclusion holds on the NUS-WIDE dataset.


\begin{table}[t]
\centering
\caption{The Training Time (s) comparing with different algorithms on both datasets.}
\label{my-label}
\scalebox{0.9}[0.9]{
\begin{tabular}{|c|c|c|c|c|c|c|}
\hline
              & \multicolumn{2}{c|}{PASCAL} & \multicolumn{2}{c|}{Wiki} & \multicolumn{2}{c|}{NUS-WIDE} \\ \hline
Methods       & 32           & 64           & 32          & 64          & 32            & 64            \\ \hline
CVH           & 5.84         & 5.18         & 0.14        & 0.04        & 3.31          & 3.25          \\ \hline
PDH           & 278.19       & 350.73       & 1.34        & 2.44        & 736.81        & 1461.71       \\ \hline
CMFH          & 324.87       & 335.41       & 0.32        & 0.59        & 1579.1        & 1782.6        \\ \hline
SMH           & 26.26        & 33.31        & 0.31        & 0.59        & 20.74         & 40.53         \\ \hline
\textbf{SMFH} & 5.89         & 6.16         & 0.56        & 0.71        & 73.94         & 88.04         \\ \hline
\end{tabular}}
\end{table}

\begin{figure}[t]
\begin{center}
\begin{minipage}[t]{0.6\linewidth}
\centerline{
\subfigure[\small{The \emph{m}AP curves vs. number of sampling pairs.}]{
\includegraphics[width=\linewidth]{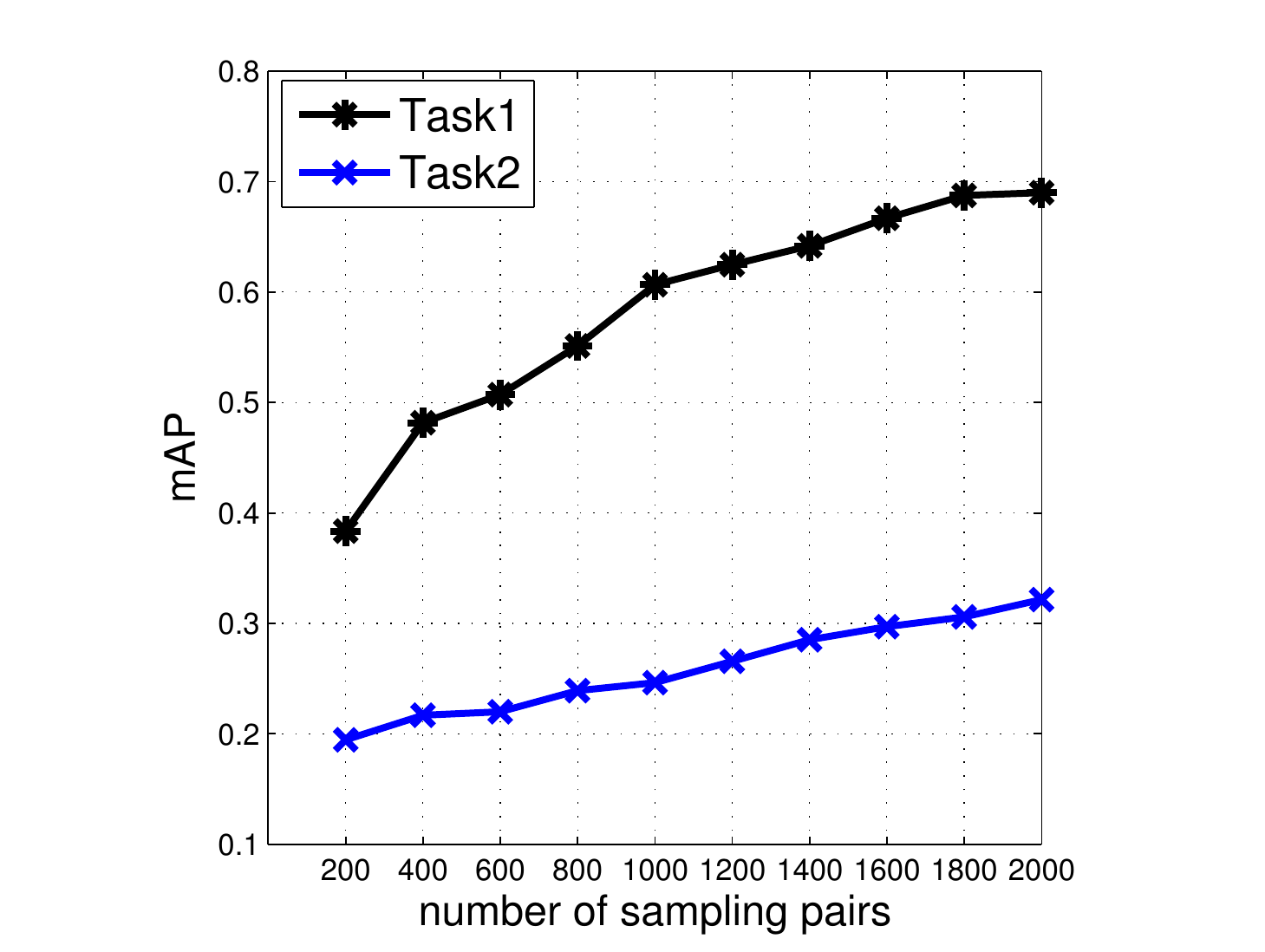}}\hspace*{-0.18\linewidth}
\subfigure[Convergence validation.]{
\includegraphics[width=\linewidth]{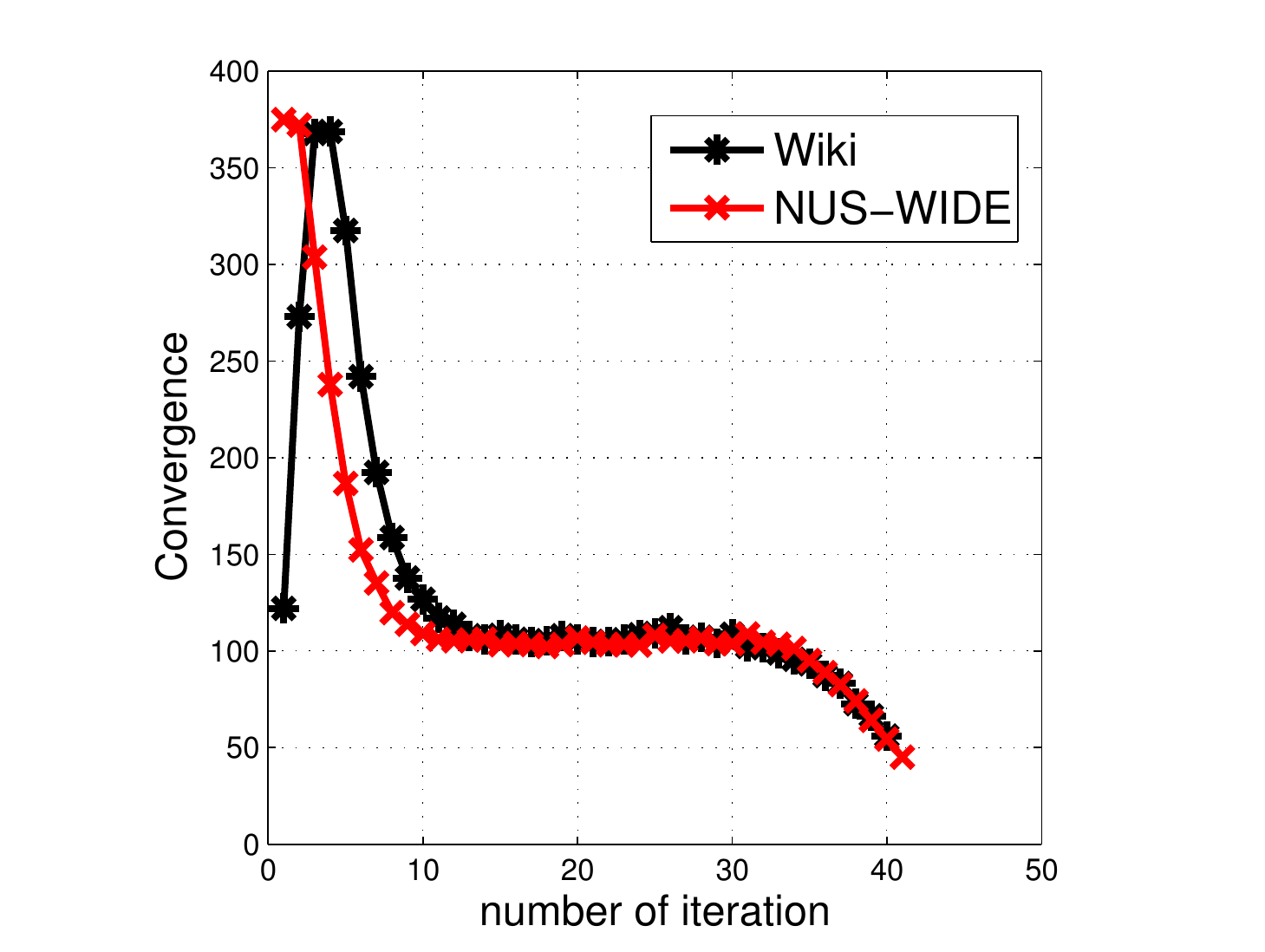}}\hspace*{-0.1\linewidth}
}
\end{minipage}
\end{center}
\setlength{\abovecaptionskip}{-5pt}
\caption{Parameter Analysis.}
\label{fig4}
\end{figure}

\section{Conclusions}\label{Sec5}
In this paper, we propose a novel hashing method termed Supervised Matrix Factorization Hashing (SMFH) for cross-modality visual search. 
We employ graph regularization to develop a collective matrix factorization based hashing framework, which can preserve the similarities among original features from different modalities into a produced Hamming space.
Meanwhile, SMFH incorporates supervised label information to enhance the quantization quality of the learned binary codes.  
Furthermore, hashing and graph regularization are integrated into a unified framework by means of joint hash function learning. 
In this framework, the given supervised labels can be leveraged to construct a label matrix, leading to more discriminative hash codes. 
Extensive experiments conducted on PASCAL-Sentence, Wiki, and NUS-WIDE benchmarks demonstrated the superior performance of SMFH over several state-of-the-art cross-modality hashing methods \cite{ding2014collective,kumar2011learning,rastegari2013predictable,zhang2014large}. 
In the future, we would investigate large-scale discrete optimization techniques for the proposed SMFH.

\section{Acknowledgement}
This work is supported by the Special Fund for Earthquake Research in the Public Interest No.201508025, the Nature Science Foundation of China (No. 61402388, No. 61422210 and No. 61373076), the Fundamental Research Funds for the Central Universities (No. 20720150080 and No.2013121026), and the CCF-Tencent Open Research Fund.

\bibliographystyle{named}
\bibliography{ijcai16}

\end{document}